\documentclass{article}
\usepackage{spconf,amsmath,graphicx, amssymb, cite, caption, booktabs, lipsum, subcaption, multirow}
\usepackage[breaklinks=true]{hyperref}
\newcommand\linesubsec[1]{\vspace{0.8mm}\noindent\textbf{#1 --- }}

\usepackage[dvipsnames]{xcolor}
\newcommand\myshade{70}
\colorlet{mywholecolor}{MidnightBlue}
\hypersetup{
  linkcolor  = mywholecolor!\myshade!black,
  citecolor  = mywholecolor!\myshade!black,
  urlcolor   = mywholecolor!\myshade!black,
  colorlinks = true,
}

\definecolor{blue}{HTML}{8ea0cb} 
\definecolor{grey}{HTML}{b3b3b3} 
\definecolor{orange}{HTML}{fc8e62} 

\title{Modelling black-box audio effects with  \\ time-varying feature modulation}
\name{Marco Comunità$^1$ \qquad Christian J. Steinmetz$^1$ \qquad Huy Phan$^{2*}$\vspace{-0.3cm}\thanks{\hspace{-16pt} \scriptsize $^{*}$Work done when H. Phan was at Centre for Digital Music, prior to joining Amazon.} \qquad Joshua D. Reiss$^1$}
\address{$^1$Centre for Digital Music, Queen Mary University of London, UK\\$^2$Amazon Alexa, Cambridge, MA, USA}
\begin{document}
\ninept
\maketitle

%===================================================%
%   ABSTRACT
%===================================================%
\begin{abstract}
Deep learning approaches for black-box modelling of audio effects have shown promise, however, the majority of existing work focuses on nonlinear effects with behaviour on relatively short time-scales, such as guitar amplifiers and distortion.
While recurrent and convolutional architectures can theoretically be extended to capture behaviour at longer time scales, we show that simply scaling the width, depth, or dilation factor of existing architectures does not result in satisfactory performance when modelling audio effects such as fuzz and dynamic range compression.
To address this, we propose the integration of time-varying feature-wise linear modulation into existing temporal convolutional backbones, an approach that enables learnable adaptation of the intermediate activations.
We demonstrate that our approach more accurately captures long-range dependencies for a range of fuzz and compressor implementations across both time and frequency domain metrics. We provide sound examples, source code, and pretrained models to faciliate reproducibility\footnote{\href{https://mcomunita.github.io/gcn-tfilm\_page}{https://mcomunita.github.io/gcn-tfilm\_page}}.
\end{abstract}


%===================================================%
%   INTRO
%===================================================%
%\vspace{-4pt}
\section{Introduction}\label{sec:intro}
\vspace{-4pt}

Audio effects are tools employed by audio engineers and musicians central to shaping the timbre, dynamics, and spatialisation of sound~\cite{wilmering2020history}. 
Digital emulation of audio effects, often referred to as virtual analogue, is an area of active research~\cite{karjalainen2006wave, yeh2009automated,  eichas2015block, eichas2017virtual, gerat2017virtual} with methods often categorised into white-, grey- and black-box approaches.
White-box modelling relies on complete knowledge of the system and often employs differential equations, which enables high quality emulations but often entails a time consuming design process and computationally expensive models~\cite{esqueda2021differentiable, parker2022physical}.
Grey-box approaches~\cite{colonel2022reverse, nercessian2021lightweight} combine a partially theoretical model with input-output measurements. This greatly reduces the prior knowledge necessary to model a device while maintaining interpretability, but still requires understanding of the underlying implementation and carefully designed measurement and optimisation procedures.
This motivates black-box models that enable emulations using only measurements from the device.
Recently, deep learning approaches have seen success in modelling a range of effects~\cite{covert2013vacuum, martinez2020deep, parker2019modelling, steinmetz2021efficient}. 
These approaches often leverage recurrent or convolutional networks operating in the time domain~\cite{damskagg2019deep, damskagg2019distortion, schmitz2018nonlinear, wright2019real}. 
While successful for some effects, modelling behaviours on longer time scales has proven challenging and is so far less investigated.%, especially when considering constraints on computation.

\begin{figure}
    \centering
    \includegraphics[width=0.9\linewidth,trim={0.2cm 0.0cm 1.12cm 0.0cm},clip]{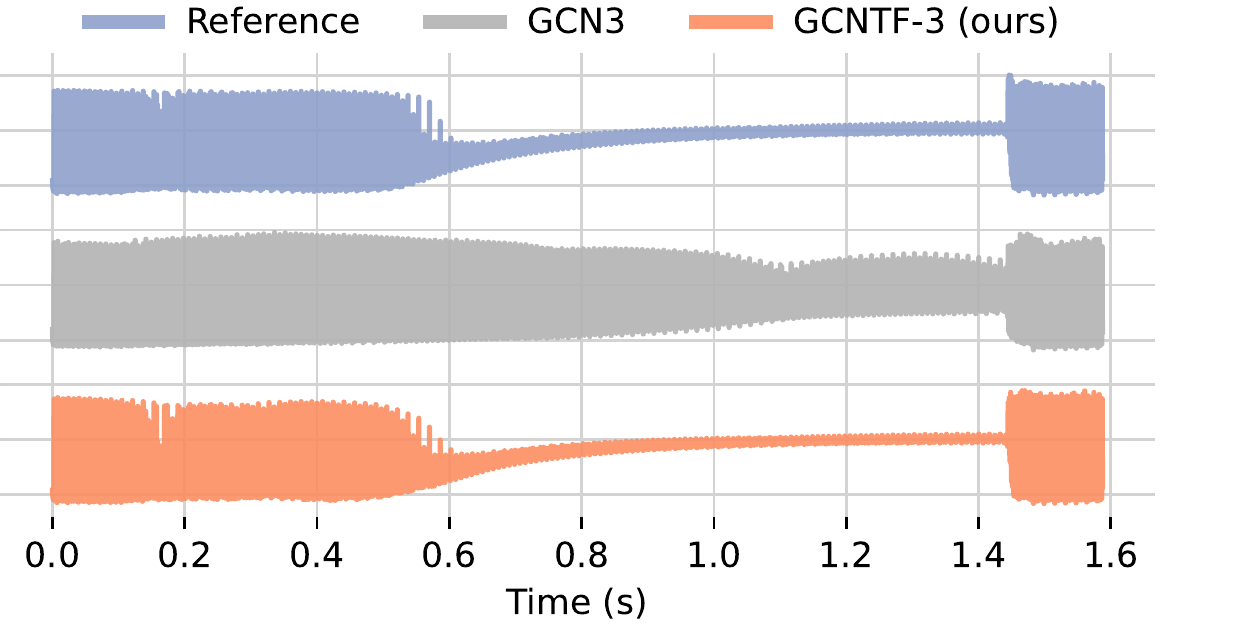}
    \vspace{-0.1cm}
    \caption{State-of-the-art black-box models like GCN-3~\cite{wright2020real} (\textcolor{grey}{grey}) fail to capture the behaviour of effects with large time constants such as fuzz~(\textcolor{blue}{blue}). Our proposed approach GCNTF-3~(\textcolor{orange}{orange}), which extends previous convolutional networks with time-varying feature modulation, enables accurate modelling of this behaviour.}
    \label{fig:waveforms}
    \vspace{-0.2cm}
\end{figure}

In this work, we focus on nonlinear time-varying audio effects that exhibit input-dependant behaviour over long time scales, such as fuzz distortion and dynamic range compression.
Distortion effects often present a challenge due to their highly nonlinear behaviour, which has been addressed in previous works relying on convolutional networks with short receptive field \cite{damskagg2019distortion} or simple one-layer recurrent networks \cite{wright2019real}.
However, distortion effects such as fuzz can also pose an additional challenge since they exhibit time-varying behaviour over larger time scales due to the attack and release of the circuit.
%Fuzz is a type of distortion that has received little attention, whilst presenting some unique challenges that, once addressed, could extend black-box modelling to a wider range of applications.
Fuzz is characterised not only by asymmetrical clipping, which for sinusoidal inputs results in a rectangular wave output, but also for its attack and release time constants which modulate the behaviour of the device as a function of the input. 
This results in a characteristic time-varying distortion, which existing deep learning based approaches fail to accurately capture, as shown in Fig.~\ref{fig:waveforms}.
%Also, the amount of duty-cycle modulation is dependent on the input level and spectrum; furthermore, attack and release times of the modulation are specific to the electronic circuit.

% and example of this is the iconic Fuzz Face\footnote{\href{https://www.electrosmash.com/fuzz-face}{https://www.electrosmash.com/fuzz-face}}. 

% extra 
%; although, attack and release are not among the controls afforded to the user.
%, dependency on the past signal, when present, is limited to tens of ms, and have been modelled in the past 

Dynamic range compressors also exhibit time-varying nonlinear behaviour over a range of timescales.
In some cases the release times of compressors can reach several seconds, such as in the classic LA-2A compressor. 
Modelling of compression has been addressed employing a range of strategies including convolutional-recurrent architectures~\cite{martinez2019general}, time-frequency representations through autoencoding~\cite{hawley2019signaltrain}, and shallow temporal convolutional networks with large receptive field~\cite{steinmetz2021efficient}.
However, performance when modelling configurations with large release time constants has not been investigated. 

We propose a method to model the nonlinear behaviour over large time-scales such as fuzz and dynamic range compression by incorporating time-varying feature modulation (Temporal FiLM)~\cite{birnbaum2019temporal} into existing temporal convolutional backbones. 
This enables adaptation of the activations of the network as a function of the input signal. 
While this is achieved through a simple mechanism of scaling and shifting the activations, we demonstrate that this enables superior performance across a range of effects without increasing the receptive field of the main network. 
Our contributions include the integration of temporal feature modulation into the black-box audio effect modelling framework and a set of benchmark datasets comprised of fuzz and compressor effects with varying time constants, which we utilise to demonstrate the failure modes of existing approaches and the ability of our proposed approach to address these limitations.
\vspace{-4pt}
\section{Method}\label{sec:method}
\vspace{-4pt}

Audio effects are signal processing devices that given an input $x \in \mathbb{R}^L$ with $L$ samples and a set of $P$ parameters $\phi \in \mathbb{R}^{P}$ that control the operation of the system, output a modified version $y \in \mathbb{R}^L$ of the signal. 
%The aim of audio effects modelling is to create a digital emulation of an analogue effect unit given measurements from this device. 
% In black-box modelling, the emulation is achieved without any specific assumption about the structure and inner workings of the unit under test. The modelling process relies solely on the input-output signals and parameters' values. 
In this work, we focus on modelling the input-output function $y=f(x,\phi)$, at one configuration of the device, holding $\phi$ constant. %Several methods are available to extend black-box models with parametric control and have been demonstrated to work~\cite{damskagg2019distortion, wright2019real, hawley2019signaltrain, steinmetz2021efficient}.
We aim to design a neural network $g_{\theta}(x)$ that produces a signal $\hat y$ perceptually indistinguishable from the real output $y$.
The modelling process involves training $g_{\theta}(x)$ with a dataset of $E$ examples ${\cal D}=\{(x_{i},y_{i},\phi)\}_{i=1}^{E}$ containing input-output recordings $(x_{i}, y_{i})$ at fixed parameters $\phi$. 
A loss function ${\mathcal L}(\hat y,y)$ measures the difference between the output of the network and the target system, providing a means to update the weights $\theta$ through a given number of gradient-based optimisation steps. 
%A successful model will accurately capture the behaviour of the system across the space of realistic input signals $\mathcal X$.

%===================================================%
\vspace{-4pt}
\subsection{Modelling Network}
\vspace{-4pt}
Similar to previous work on distortion effect modelling~\cite{damskagg2019distortion, martinez2020deep}, we adopt a feedforward WaveNet \cite{rethage2018wavenet} architecture, also known as a temporal convolutional network (TCN). 
We refer to this architecture as the Gated Convolution Network (GCN) since it a special case of the TCN that utilises gated convolutions. 
The GCN is composed of $M$ blocks with each block containing $\frac{N}{M}$ layers for a total of $N$ layers. 
Each layer in a block is made of a dilated 1-dimensional convolutional layer followed by a gated activation as shown in Fig.~\ref{fig:blocks} (left). 
The outputs from each layer are summed through a $1 \times 1$ convolution to produce the final output $y$. 
We implement several variants of the base GCN architecture with short, medium and long receptive fields relying on different number of layers and kernel sizes, and make also use of rapidly growing dilation factors~\cite{steinmetz2021efficient, tian2020tfgan, yang2020multi}. 

%===================================================%
\vspace{-4pt}
\subsection{Temporal FiLM}\label{sec:tfilm}
\vspace{-4pt}
 Feature-wise Linear Modulation (FiLM) is a general-purpose conditioning method that operates on the intermediate features of a neural network as a function of conditioning signals~\cite{perez2018film}. 
Given a conditioning signal $
\mathbf{x}_{i}$, FiLM learns two functions $f$ and $g$, which are used to map the conditioning signal to a set of scaling $\gamma_{n,c} = f(\mathbf{x}_{i})$ and bias $\beta_{n,c} = h(\mathbf{x}_{i})$ parameters for each layer $n$ and channel $c$ of the network. 
These parameters are used to modulate the intermediate activations at each layer $\mathbf{z}_{n,c}$ via a feature-wise affine transformation
\begin{equation}
\text{FiLM}(\mathbf{z}_{n,c}, \gamma_{n,c}, \beta_{n,c})=\gamma_{n,c} \cdot \mathbf{z}_{n,c} + \beta_{n,c}.
\end{equation}
In practice, $f$ and $h$ are implemented as a neural network and can be learned during training of the main network.

While FiLM has proven to be a powerful conditioning method, it can be further extended to increase the expressivity of the network by leveraging long-range dependencies in the conditioning signal to vary the modulation of intermediate features across time; an approach known as Temporal Feature-wise Linear Modulation (TFiLM)~\cite{birnbaum2019temporal}. 
Using recurrent networks, TFiLM layers modulate the intermediate features of a convolutional model over time as a function of the activations at each layer.
This has conceptual connections to other input dependant and time-varying conditioning approaches such as hypernetworks~\cite{ha2016hypernetworks} and dynamic convolution~\cite{chen2020dynamic}, which enable adaptation of the weights of convolutional networks.
However, TFiLM provides a simpler method for adaptation that is both efficient and often easier to train.
Thus far, TFiLM has only been applied to the task of audio super resolution using UNet-like architectures and has not yet been integrated in the GCN/TCN architecture, as we propose in this work. 

\begin{figure}[t]
    \centering
    \includegraphics[width=0.95\linewidth, trim={0.0cm 0.2cm 0.8cm 0.1cm},clip]{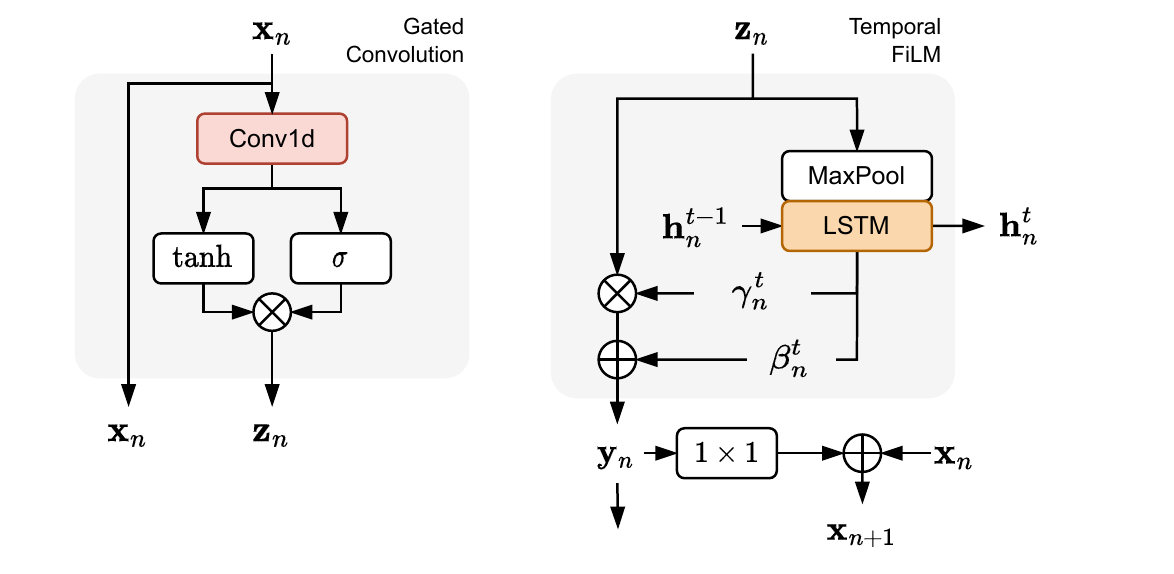}
    \vspace{-0.2cm}
    \caption{Block diagram of the dilated 1-dimensional gated convolution block (left) and the Temporal FiLM module (right).}\vspace{0.3cm}
    \label{fig:blocks}
\end{figure}
\begin{figure}[t]
    \centering
    \includegraphics[width=1.0\linewidth, trim={0.8cm 0.0cm 1.2cm 0.1cm},clip]{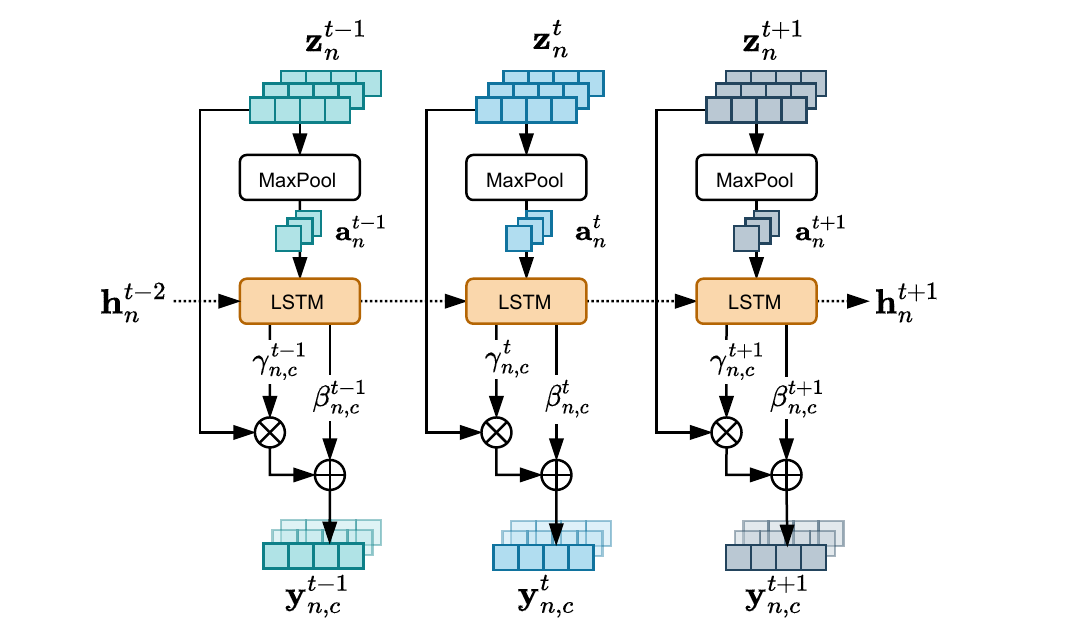}
    \vspace{-0.3cm}
    \caption{Temporal FiLM modulates the intermediate activations of a convolutional network at each layer by splitting feature maps along the sequence dimension into $T$ blocks $\mathbf{z}_{n}^t$. Max pooling is applied to each block to generate $\mathbf{a}_{n}^t$, which is used as input to the LSTM (with hidden activations $\textbf{h}^t_n$) that generates scaling and bias parameters. This illustrates a case when $B=4$ and $C=3$. For clarity only the affine transformation of the first channel is shown.}
    \label{fig:tfilm}
\end{figure}
%\vspace{1.6cm}

To capture both nonlinear behaviour and long range temporal dependencies in modelling audio effects, we propose to integrate time-varying feature-wise linear modulation in the base GCN architecture, which we refer to as GCN with Temporal FiLM (GCNTF).
In our formulation, given a sequence of activations $\mathbf{z}_{n} \in \mathbb{R}^{C \times L}$ from the $n$-th layer of a GCN, where $C$ is the number of channels, and $L$ is the sequence dimension, we split the sequence into $T$ blocks of $B$ samples along the sequence dimension. 
For each block $\mathbf{z}^{t}_{n}$, 1-dimensional max pooling is applied to downsample the signal by a factor of $B$ to produce $\mathbf{a}^{t}_{n}$ as shown in Fig.~\ref{fig:tfilm}.
Then an LSTM generates a sequence of scaling and bias parameters $\text{LSTM}_n(\mathbf{a}^{t}_{n}) =   (\gamma^{t}_{n,0}, ..., \gamma^{t}_{n,c}), (\beta^{t}_{n,0}, , ..., \beta^{t}_{n,c}$) for each channel $c$.
These scaling and bias parameters are then used to modulate each channel of the activations individually in each block by an affine transformation
\begin{equation}
 \mathbf{y}^{t}_{n,c} = \gamma^{t}_{n,c} \mathbf{z}^{t}_{n,c} + \beta^{t}_{n,c}. 
\end{equation}

As shown in Fig.~\ref{fig:blocks} (right), the output of each TFiLM module is sent through a $1 \times 1$ convolution and combined with the residual connection $\mathbf{x}_n$ and sent to the following layer. 
The output of the TFiLM module $y_n$ is sent to the final layer of the network where all intermediate outputs are mixed together via another $1 \times 1$ convolution. 
%The activations $Z_{nm}$ are then modulated by the output of a TFiLM layer and the modulated sequence $Z_{nm}^{'}$ sent to the final mixing layer. 
%Each sequence $Z^{'}_{nm}$ is also mixed with the input $X_{nm}$ via a residual connection $X_{n,m+1}$ and sent to the following layer.
%Like in the original work, we use max pooling to downsample the sequence by a factor of B, and LSTM layers to generate the modulation sequences.

%\begin{figure}
%\centering
%\begin{subfigure}{.45\linewidth}
% \centering
%\includegraphics[width=\linewidth]{figures/GCN Conv.pdf}
%\caption{GCNTF layer}
%\label{fig:gcn_conv}
%\end{subfigure}%
%\begin{subfigure}{.5\linewidth}
%    \centering
%    \includegraphics[width=\linewidth]{figures/TFiLM.pdf}
%    \caption{Temporal FiLM layer}
%    \label{fig:tfilm}
%\end{subfigure}
%\vspace{0.3cm}
%\caption{GCN architecture with temporal FiLM.}
%\label{fig:test}
%d\end{figure}

%===================================================%
%\vspace{-4pt}

%\vspace{-4pt}

%===================================================%
%   Experiments
%===================================================%
%\newpage
\vspace{-4pt}
\section{Experimental Design}\label{sec:exp}
\vspace{-4pt}
To understand how Temporal FiLM aids in the modelling process, specifically for effects with behaviour at long time scales, we train models across two effect classes: fuzz and compressor. % , using varying release time constants at 50, 250, and 2500\,ms
As baselines, we considered state-of-the-art convolutional networks~\cite{damskagg2019deep, damskagg2019distortion, wright2020real} proposed for guitar amplifier and distortion effect modelling. 
The models (GCN-1 and GCN-3) have respectively, 1 block of 10 layers and 2 blocks of 9 layers, dilation growth of 2, kernel size of 3 and 16 channels for every convolutional layer; giving receptive fields of 2047 and 2045 samples ($\approx45$\,ms at $f_s= 44.1$\,kHz). 

While GCN-1 and GCN-3 were successful in modelling guitar amplifiers and distortion effects, they may not be capable of modelling effects with longer temporal behaviour due to their relatively small receptive field. 
To address this, we construct stronger baselines by extending these models to create variants with longer receptive fields by adopting larger dilation growth~\cite{steinmetz2021efficient}.
As a result, GCN-250 has 1 block of 4 layers, kernel size of 41, and dilation growth of 6, for a receptive field of 250\,ms, while GCN-2500 has 1 block of 10 layers, kernel size of 5 and dilation growth of 3 for a receptive field of 2500\,ms.
As further baselines, we also include state-of-the-art recurrent networks (LSTM-32 and LSTM-96)~\cite{wright2020real}.
% We modify all these models using our proposed method and - adding TFiLM layers as introduced in previous sections - we obtain similar architectures that also include time-varying feature modulation.
To validate our approach we then added TFiLM layers to each of the baseline models to enable time-varying feature modulation.
We refer to these models as: GCNTF-1, GCNTF-3, GCNTF-250, GCNTF-2500, which results in a total of 12 different models that we considered.

%===================================================%
\vspace{-4pt}
\subsection{Experiments}
\vspace{-4pt}

\linesubsec{Time constants}
To evaluate the ability of the models to capture behaviour over long time scales, we created a set of specialised datasets for fuzz and compressor effects. 
For fuzz, we designed an analogue circuit (Custom Fuzz) which includes, together with the typical volume and gain, attack and release controls, which was designed using LiveSpice\footnote{\href{https://www.livespice.org}{https://www.livespice.org}}.
For the compressor, we used the implementation in the Pedalboard library\footnote{\href{https://github.com/spotify/pedalboard}{https://github.com/spotify/pedalboard}}, which enables arbitrary control over the time constants.
%, which also includes a VST host plugin to run the simulation within a digital audio workstation.
With these implementations, we then assembled a dataset of processed electric guitar signals using recordings from a subset of the IDMT-SMT-Guitar dataset~\cite{kehling2014automatic}, which contains short musical pieces recorded with two different guitars, for a total of $\approx28$\,min of audio. 
Clean and processed audio were split in $\approx14$\,min training data and $\approx7$\,min each for validation and test.
% Clean and processed audio were split in training, validation and test set of about 14, 7 and 7 minutes.
To make sure to capture the complex dynamic behaviour of our design the input signal's amplitude changes randomly every 5 sec.
For Fuzz, each model was trained with 3 different settings for attack and release times, respectively: 50\,ms and 50\,ms, 10\,ms and 250\,ms, 1\,ms and 2500\,ms, while for compressor attack and release were set to, respectively: 10\,ms and 50\,ms, 5\,ms and 250\,ms, 1\,ms and 2500\,ms.
%We repeat similar experiments on a basic digital compressor from the Spotify Pedalboard library\footnote{\href{https://github.com/spotify/pedalboard}{https://github.com/spotify/pedalboard}}, which allows to choose any value for the two time constants. In this case attack and release were set to, respectively: 10ms and 50ms, 5ms and 250ms, 1ms and 2500ms.

%\begin{figure}
%    \centering
%    \includegraphics[width=\linewidth]{figures/Fuzz block diagram.pdf}
%    \caption{Block diagram of our Custom Fuzz design}
%    \label{fig:fuzz_block}
%\end{figure}

%===================================================%
\linesubsec{Other effects}
To further demonstrate the performance of our approach we also trained models on the Fuzz Face emulation plugin from Distorque Audio\footnote{\href{http://distorqueaudio.com/plugins/face-bender.html}{http://distorqueaudio.com/plugins/face-bender.html}}, the LA-2A compressor from the SignalTrain dataset~\cite{hawley2019signaltrain}, and digital compressor, MCompressor, by Melda Production\footnote{\href{https://www.meldaproduction.com/MCompressor}{https://www.meldaproduction.com/MCompressor}}. 
For the MCompressor, we selected two attack and release settings: 5\,ms and 250\,ms, 1\,ms and 1000\,ms. 
The LA-2A has no attack and release controls, but it is a complex analogue compressor design with a ``\textit{ratio of 3:1, a frequency dependent average attack time of 10\,ms and a release time of about 60\,ms for 50\% of the release, and anywhere from 1 to 15 seconds for the rest}''\footnote{\href{https://www.uaudio.com/blog/la-2a-collection-tips-tricks/}{https://www.uaudio.com/blog/la-2a-collection-tips-tricks/}}.

%===================================================%
\linesubsec{Channel width}
To ensure performance of models with TFiLM is not simply due to the increase in number of trainable parameters, we compared GCNTF-3 with a larger variant of the GCN-3 model with $C=24$ channels. This results in both models having $\sim71$k parameters. These models were then trained on the Custom Fuzz with 1\,ms attack and 2500\,ms release.

%===================================================%
\linesubsec{Block size}
We conclude our experiments measuring the performance of model with TFiLM as a function of the block size $B$, which relates to the downsampling factor and adaptation rate. We ran experiments with block sizes of $B \in 32, 64, 128, 256, 512$, training GCNTF-3 on the Custom Fuzz with 1\,ms attack and 2500\,ms release.

%===================================================%
\vspace{-4pt}
\subsection{Training details}
\vspace{-4pt}
All models were trained with Adam with weight decay of $1 \cdot 10^{-4}$ and an initial learning rate of $5 \cdot 10^{-3}$. 
The learning rate was halved whenever the validation loss saw no improvement for 10 epochs. 
We used early stopping with a patience of 40 epochs on the validation loss and limited training to 2000 epochs, with most models training for less than 400 epochs and none reaching the limit. 
All models were trained at $f_s = 44.1$\,kHz with inputs of 112640 samples ($\approx 2.5$\,s) and batch size of 6.
%All models were trained with a single GPU for 
%\subsection{Loss function}\label{sec:loss}
% We used a combination of the error in the time and frequency domains. We compute the mean absolute error (MAE) for the time domain component ${\cal L}_{time}$ and the multi-resolution short-time Fourier Transform error~\cite{steinmetz2020auraloss, yamamoto2020parallel} for the frequency domain ${\cal L}_{freq}$ component as used in previous work \cite{steinmetz2020mixing}. 
We used a combination of the error in the time and frequency domains, respectively: mean absolute error (MAE) and multi-resolution short-time Fourier Transform error (MR-STFT)~\cite{steinmetz2020auraloss, yamamoto2020parallel}, as in previous work~\cite{steinmetz2020mixing, steinmetz2021efficient}. 
The overall loss is a sum of two terms ${\cal L} = {\cal L}_{\text{MAE}} + \alpha {\cal L}_{\text{MR-STFT}}$, with $\alpha = 1$.

\begin{figure}[]
    \centering
    \includegraphics[width=0.9\linewidth, trim={0.0cm 0.5cm 0.0cm 1.0cm},clip]{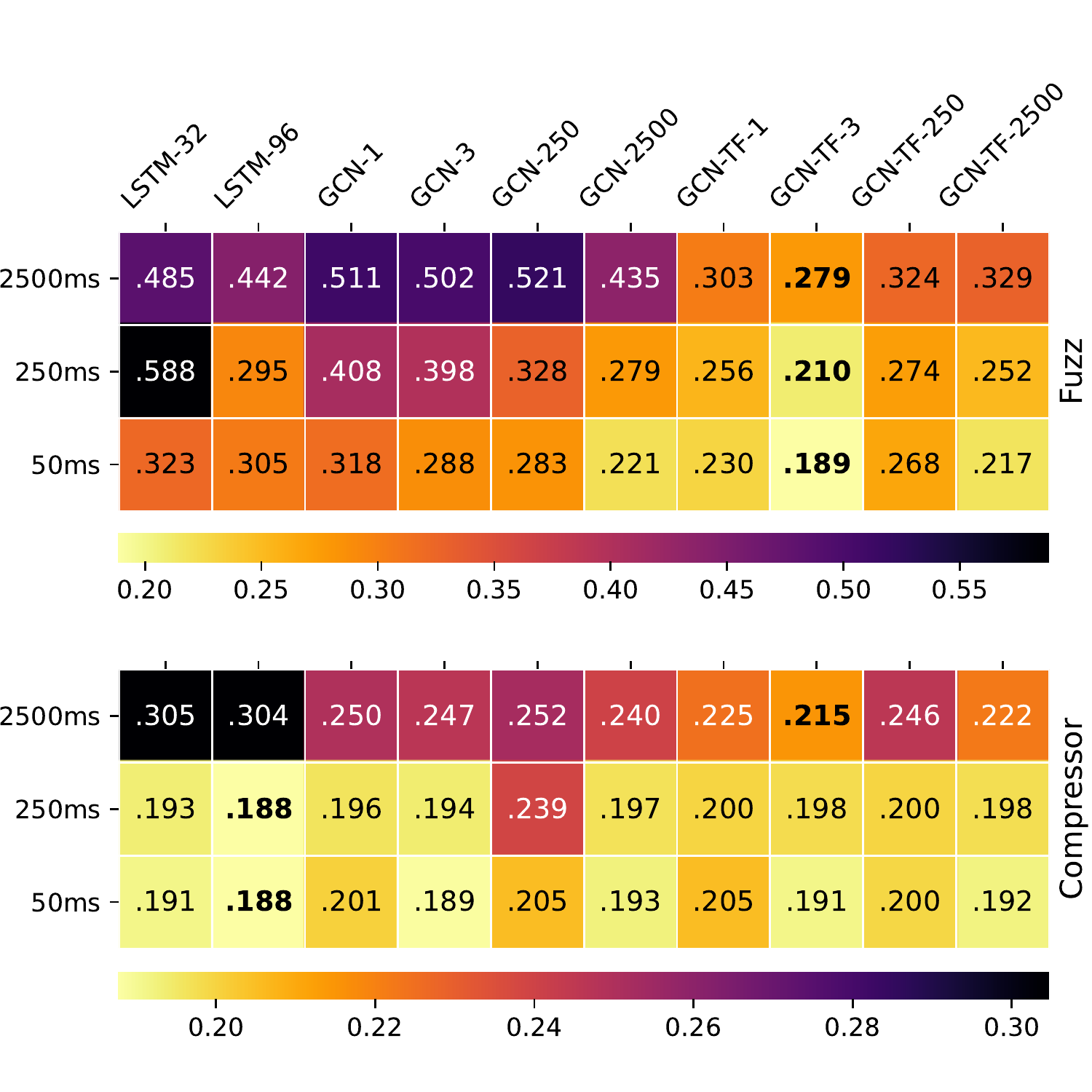}
    % \vspace{-12pt}
    \vspace{-0.2cm}
    \caption{MR-STFT error for models of fuzz (Custom Fuzz) and compressor (Pedalboard) effects with varying time constants. The best performing models for each effect configurations are in boldface.}
    \label{fig:stft_grid}
\end{figure}

\vspace{-0.3cm}
\renewcommand{\arraystretch}{0.85}
\begin{table*}[h]
    \centering
    \begin{tabular}{l c c c c c c c c c c} \toprule
    
        \multirow{2}{*}{Model}    & \multirow{2}{*}{Params.}   &  \multicolumn{2}{c}{LA-2A} & \multicolumn{2}{c}{MComp. (5ms/250ms)} &  \multicolumn{2}{c}{MComp. (1ms/1000ms)} & \multicolumn{2}{c}{Face Bender} \\ \cmidrule(lr){3-4} \cmidrule(lr){5-6} \cmidrule(lr){7-8} \cmidrule(lr){9-10}
                            &          & $L1$              & MR-STFT   & $L1$      & MR-STFT   & $L1$      & MR-STFT   & $L1$  & MR-STFT     \\ \midrule
        LSTM-32             
         & 4.5k  & 0.012 & 0.356 & 0.001 & 0.239 & 0.002 & 0.250 & 0.004 & 0.236 \\ 
        LSTM-96             
         & 38.1k & 0.012 & 0.323 & 0.002 & 0.275 & 0.002 & 0.278 & 0.026 & 0.379\\ \midrule
        GCN-1               
            & 17.1k & 0.012 & 0.333 & 0.001 & 0.201 & 0.002 & 0.246 & 0.004 & 0.204 \\ 
        GCN-3               
            & 32.0k & 0.001 & 0.331 & 0.022 & 0.197 & 0.002 & 0.255 & 0.002 & 0.192 \\ 
        GCN-250             
           & 65.6k & 0.002 & 0.339 & 0.022 & 0.218 & 0.033 & 0.206 & 0.004 & 0.222 \\ 
        GCN-2500            
            & 26.4k & 0.001 & 0.310 & 0.022 & 0.186 & 0.033 & 0.184 & 0.226 & 0.239 \\ \midrule
        GCNTF-1 (ours)      
            & 38.9k & 0.012 & 0.306 & 0.022 & 0.195 & \textbf{0.001} & \textbf{0.176} & 0.004 & 0.224\\ 
        GCNTF-3 (ours)      
            & 71.1k & 0.001 & 0.302 & 0.022 & 0.182 & 0.001 & 0.191 & \textbf{0.001} & \textbf{0.164} \\ 
        GCNTF-250 (ours)    
            & 74.3k & 0.011 & 0.346 & \textbf{3.0e-4} & \textbf{0.174} & 0.033 & 0.183 & 0.003 & 0.192 \\ 
        GCNTF-2500 (ours)   
            & 48.2k & \textbf{0.001} & \textbf{0.296} & 3.1e-4 & 0.179 & 0.033 & 0.167 & 0.225 & 0.213 \\ 
         \bottomrule 
    \end{tabular}
    \vspace{-0.0cm}
    \caption{Impact of using TFiLM to model LA-2A, MCompressor and Face Bender. Lowest total error for each configuration in boldface.}
    \label{tab:other_fx} \vspace{0.2cm}
\end{table*}

%===================================================%
%   RESULTS
%===================================================%
%\vspace{-4pt}
\vspace{0.4cm}
\section{Results}\label{sec:results}
\vspace{-2pt}

%\setlength{\tabcolsep}{3.8pt}
%\begin{table}[]
%    \centering
%    \begin{tabular}{l c c c c c c c c c} \toprule
%    Model   & Params. M & R.f. (ms) & \multicolumn{2}{c}{Fuzz} & \multicolumn{2}{c}{Compressor} \\ \cmidrule(lr){4-5} \cmidrule(lr){6-7}
%            &           &           & $L1$ & STFT & $L1$ & STFT  \\ \midrule
%    LSTM32 & & - & 0.000 & 0.000 & 0.000 & 0.000 \\
%    LSTM96 & & - & 0.000 & 0.000 & 0.000 & 0.000 \\ \midrule
%    GCN1  & & 50 & 0.000 & 0.000 & 0.000 & 0.000 \\
%    GCN3  & & 40 & 0.000 & 0.000 & 0.000 & 0.000 \\
%    GCN250  & & 250 & 0.000 & 0.000 & 0.000 & 0.000 \\
%    GCN2500 & & 2500 & 0.000 & 0.000 & 0.000 & 0.000 \\ \midrule
%    GCNTF1  & & 50 & 0.000 & 0.000 & 0.000 & 0.000 \\
%    GCNTF3  & & 40 & 0.000 & 0.000 & 0.000 & 0.000 \\
%    GCNTF250  & & 250 & 0.000 & 0.000 & 0.000 & 0.000 \\
%    GCNTF2500  & & 2500 & 0.000 & 0.000 & 0.000 & 0.000 \\ \bottomrule
%    \end{tabular}
%    \caption{Modelling fuzz and compressor with release time constant of 2500 ms.}
%    \label{tab:my_label}
%\end{table}

\begin{figure}[t]
    \centering
    \includegraphics[width=\linewidth]{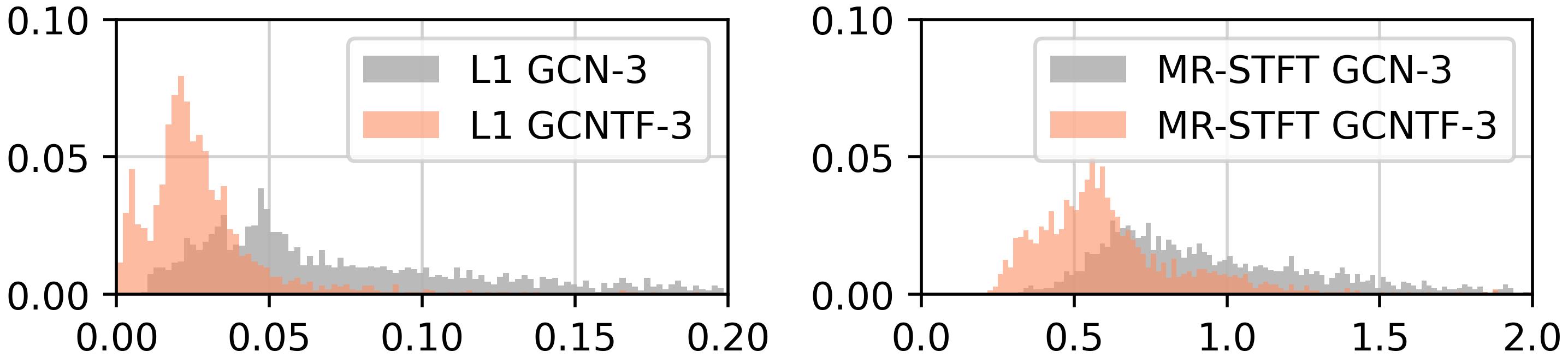}
            \vspace{-0.5cm}
    \caption{GCN-3 and GCNTF-3 modelling fuzz with 2500\,ms release \vspace{-0.0cm}}
    \label{fig:fuzz_hist}\vspace{0.3cm}
\end{figure}

\begin{figure}[t]
    \centering
    \includegraphics[width=\linewidth]{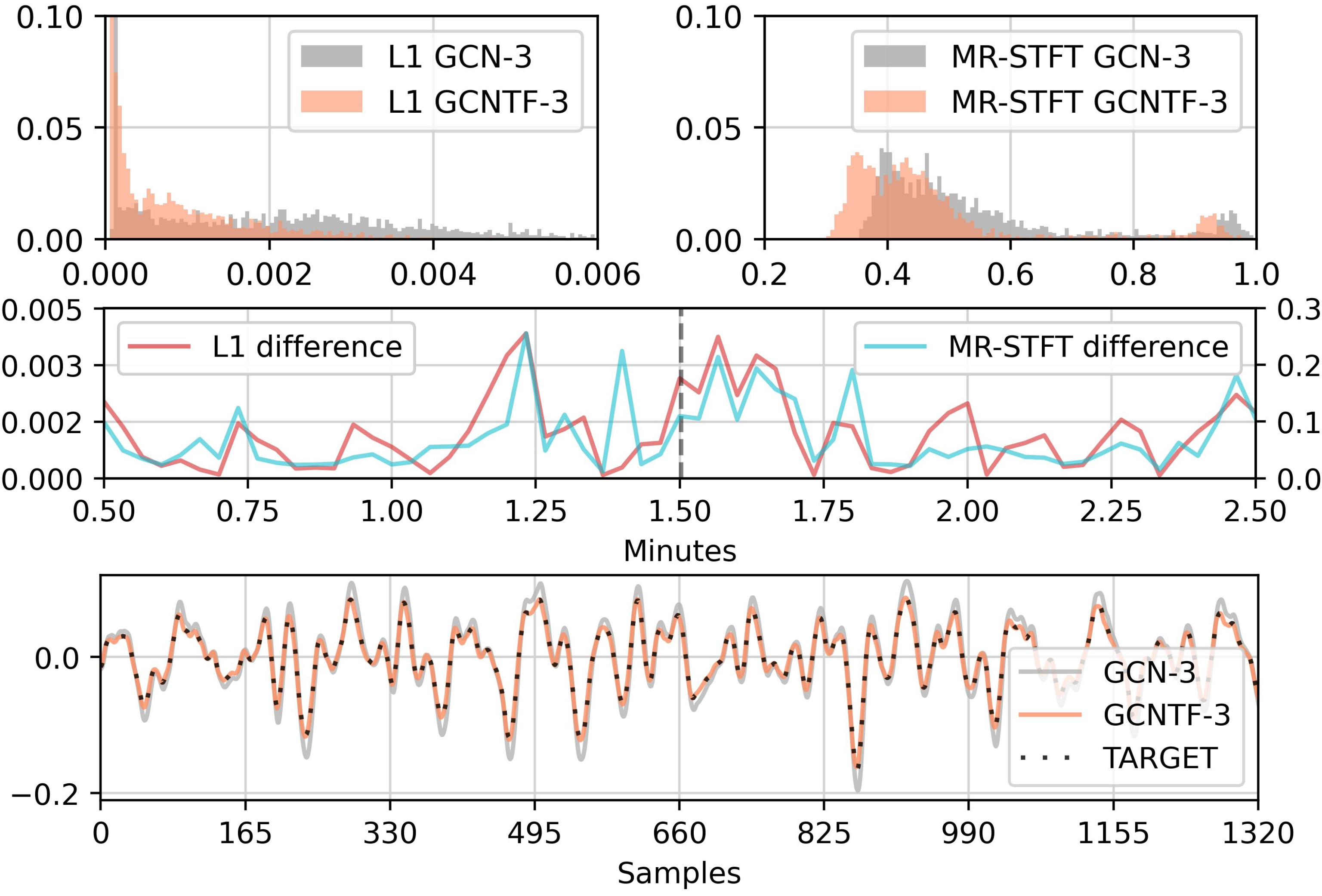}
    \vspace{-0.5cm} 
    \caption{GCN-3 and GCNTF-3 for compressor with 2500\,ms release. Top: histogram of L1 and MR-STFT error. Middle: difference between error for the two models. Bottom: waveforms at 1.50 min}
    \label{fig:comp2500}
\end{figure}

%===================================================%
\linesubsec{Time constants}
Results comparing our proposed approach with the state-of-the-art for the tasks of modelling Custom Fuzz and Pedalboard compressor across attack and release settings are shown in Fig.~\ref{fig:stft_grid}. 
%We report the MR-STFT loss for clarity and since the L1 loss had negligible impact on the total. 
Results on fuzz are consistent, with GCNTF-3 performing best regardless of the time constants and error approximately halved with respect to the GCN-3. 
This result demonstrates how the very short receptive field of the two models (2045 samples) captures the rich timbre of the fuzz on a short-range, while TFiLM enables modelling long-range dependencies with the past input. 
Furthermore, even models with a receptive field sufficient to capture the past context do not capture the device behaviour, fully motivating models with TFiLM.
To further compare the two models we propose to compute the error across time (e.g., every 8192 samples) and compare the distributions, as we do in Fig. \ref{fig:fuzz_hist} for fuzz with 2500ms release. We observe the error for GCN-3 has a higher median and heavier tails when compared to GCNTF-3. 
% Also, by looking at the difference between the losses across time it is possible to identify when
For compressor modelling the results show improvements when using TFiLM only for long release values, with GCNTF-3 performing best. 
Conversely, for both 50\,ms and 250\,ms release time, LSTM96 shows the lowest error.
To understand how the models performance differ, we show the error distribution in Fig. \ref{fig:comp2500}.
Also, by looking at the difference between L1 and MR-STFT error across time it is possible to identify a divergence around 1.50 minutes, of which we show an excerpt. 

% To further compare the two models we propose to compute the loss across time (e.g., every 8192 samples) and compare the distributions, as we do in Fig.\ref{fig:fuzz_hist} for fuzz with 2500ms release. In this way we can observe how the losses for GCN-3 have higher median and heavier tails when compared to GCNTF-3.

\setlength{\tabcolsep}{3.8pt}
\begin{table}[]
    \centering
    \begin{tabular}{l c c c c c} \toprule
        Model          & $C$    & Params.   & $L1$              & MR-STFT \\ \midrule
        GCN-3          & 16     & 31.97k    & 0.045             & 0.502 \\
        GCN-3          & 24     & 70.99k    & 0.046             & 0.505 \\ \midrule
        GCNTF-3 (ours) & 16     & 71.14k    & \textbf{0.011}    & \textbf{0.279} \\ 
        \bottomrule 
    \end{tabular}
    \vspace{-0.0cm}
    \caption{Impact of channel width $C$ in modelling the Custom Fuzz with attack of 1ms and release of 2500\,ms.}
    \label{tab:channel-width} \vspace{0.3cm}
\end{table}
\begin{table}[]
    \centering
    \begin{tabular}{l c c c} \toprule
        Model           & $B$   & $L1$              & MR-STFT \\ \midrule
        GCN-3           & -     & 0.045             & 0.502 \\ \midrule
        GCNTF-3 (ours)  & 32    & 0.042             & 0.538 \\ 
        GCNTF-3 (ours)  & 64    & 0.013             & 0.307 \\ 
        GCNTF-3 (ours)  & 128   & \textbf{0.011}    & \textbf{0.279} \\
        GCNTF-3 (ours)  & 256   & 0.013             & 0.288 \\ 
        GCNTF-3 (ours)  & 512   & 0.015             & 0.314 \\ 
         \bottomrule 
    \end{tabular}
    \vspace{-0.0cm}
    \caption{Impact of feature modulation block-size $B$ in modelling the Custom Fuzz with attack of 1\,ms and release of 2500\,ms.}
    \label{tab:block-size}
\end{table}

%===================================================%
\linesubsec{Other effects}
% To verify the performance of our proposed method for complex and renowned compressor and fuzz designs we also considered popular implementations of these effects. 
Results for Face Bender, LA-2A, and MCompressor are shown in Table~\ref{tab:other_fx}. In almost all cases the use of time-varying feature modulation results in lower overall error, showing a generalised improvement with respect to the state-of-the-art. 

%===================================================%
\linesubsec{Channel width}
% In Table \ref{tab:channel-width} we report the results of our experiments on models with equal number of parameters, showing how the performance improvement of models with TFiLM layers is not simply due to the increase of trainable parameters.
Table~\ref{tab:channel-width} demonstrates how increasing the number of the parameters of the base GCN model does not lead to a performance improvement. This indicates that is not simply the increased number of parameters in TFiLM models that leads to an improvement, but instead time-varying modulation of activations.

%===================================================%
\linesubsec{Block size}
% We also analyse the results for our proposed method at different block sizes $B$ and report the results in Table \ref{tab:block-size}.
Results for our proposed method at different block sizes are reported in Table \ref{tab:block-size}.
For GCNTF-3 trained on Custom Fuzz, there seems to be an optimal block size of $B=128$ samples.
\vspace{-4pt}
\section{Conclusion}\label{sec:Conclusion}
\vspace{-4pt}
In this work, we presented a method for black-box modelling of audio effects with long-range dependencies by integrating time-varying feature-wise modulation into state-of-the-art convolutional models.
We demonstrated that current state-of-the-art approaches fail to model behaviours over long time-scales for effects like fuzz and compressor, while our proposed method successfully captures them without increasing the receptive field of the processing network.
These results open up future work to extend the approach to time-varying effects like chorus or tremolo, but also in applications like the proxy network approach for learning to control effects.
Datasets, source code, and pretrained models are openly provided.

%===================================================%
%   ACKNOWLEDGEMENTS
%===================================================%
\vspace{-4pt}
\section{Acknowledgements}\label{sec:acknowledgement}
\vspace{-4pt}
Funded by UKRI and EPSRC as part of the ``UKRI CDT in Artificial Intelligence and Music'', under grant EP/S022694/1.

% \vfill\pagebreak

%===================================================%
%   REFERENCES
%\clearpage
\bibliographystyle{IEEEbib}
\bibliography{refs}

\end{document}